\documentclass[sigconf, screen]{acmart}

\AtBeginDocument{%
  }

\setcopyright{none} 
\copyrightyear{2026}
\acmYear{2026}
\acmDOI{XXXXXXX.XXXXXXX}

\usepackage{xcolor}
\usepackage{amsmath}
\usepackage{cleveref}
\usepackage{amsfonts}
\usepackage{stmaryrd}
\usepackage{marvosym}
\usepackage{lipsum}

\usepackage{makecell}
\usepackage[normalem]{ulem}
\usepackage{enumitem}

\newcommand{\change}[1]{{#1}}
\usepackage{algorithm}
\usepackage{algorithmic}
\usepackage{pifont}
\usepackage{multirow}
\usepackage{subcaption}
\usepackage{float}
\usepackage{svg}

\usepackage{setspace}

\usepackage{graphicx}
\usepackage{upgreek}

\begin{document}

\title{Trusted Hardware Acceleration for Function Secret Sharing}
\author{Pengzhi Huang}
\affiliation{%
  \institution{Cornell University}
  \city{Ithaca}
  \state{NY}
  \country{USA}}
\email{ph448@cornell.edu}

\author{Kiwan Maeng}
\affiliation{%
  \institution{Pennsylvania State University}
  \city{University Park}
  \state{PA}
  \country{USA}}
\email{kvm6242@psu.edu}

\author{G. Edward Suh}
\affiliation{%
  \institution{NVIDIA / Cornell University}
  \city{Westford}
  \state{MA}
  \country{USA}}
\email{esuh@nvidia.com}




\begin{abstract}
Function secret sharing (FSS) is a core building block for privacy-preserving systems such as secure inference and private information retrieval (PIR), but incurs significant overhead in key generation, communication, and data movement.
We present the distributed function accelerator (DFA), a hardware accelerator that targets the dominant primitive in FSS: distributed point function (DPF) generation and evaluation. DFA combines a high-throughput fixed-function engine for AES-based pseudorandom number generation with a lightweight programmable unit for protocol-specific logic. In untrusted mode, DFA serves as a pure accelerator for DPF evaluation, improving throughput and energy efficiency without changing the protocol. In trusted mode, it further enables local, on-the-fly key generation, eliminating key distribution, and reducing storage and data movement overheads.
Across representative workloads, DFA achieves a reduction of $10\times$ end-to-end latency, a reduction of up to $20\times$ communication and more than $5\times$ energy savings for secure inference; and an improvement of $5\times$ throughput and $10\times$ energy reduction for PIR, with modest hardware cost.
\end{abstract}

\maketitle

\section{Introduction}
As machine learning (ML) becomes central to applications ranging from healthcare diagnostics~\cite{kaissis2021end,wong2020deep,yang2017big} to cloud-based analytics services~\cite{Azure,GoogleAI}, these systems increasingly rely on processing sensitive data such as patient records, proprietary models, and personal identifiers. In many real-world deployments, data are distributed across institutions or even nations, or are submitted by users to cloud-hosted models, raising significant privacy risks that cannot be fully mitigated by policy or regulation alone. Cryptographic techniques for privacy-preserving computation---such as secure multi-party computation (MPC) for private ML inference~\cite{yao1982protocols,wagh2019securenn,rathee2020cryptflow2} and private information retrieval (PIR)~\cite{chor1998private}---offer strong confidentiality guarantees by enabling computation over protected data.
Despite recent advances that make these approaches increasingly practical~\cite{rathee2020cryptflow2,wagh2020falcon,angel2018pir,mohassel2018aby3,demmler2015aby}, they remain bottlenecked by substantial system-level overheads, including heavy communication overhead in MPC~\cite{mohassel2018aby3,crypten2020,kumar2020cryptflow,wagh2020falcon}, high computational and memory costs in fully homomorphic encryption~\cite{reagen2021cheetah,kim2024cheddar}, and data movement and storage challenges in large-scale deployments~\cite{gupta2023sigma}. 

Function secret sharing (FSS)~\cite{boyle2016function} has emerged as a promising building block for both MPC-based secure inference and PIR.
When using FSS for MPC-based secure inference (i.e., \emph{FSS-based} private inference), most of the computation and communication overhead is shifted to the offline key generation phase, making the online phase relatively lightweight and high throughput~\cite{boyle2016function,de2022lightweight, gupta2023sigma,jawalkar2024orca}. 
FSS can also be a key building block for multi-party private information retrieval (PIR), which is used for private data lookup of a remote database without revealing the index of accessed data to the server~\cite{wang2017splinter,lam2023gpu,xie2024access}.
Other privacy-preserving applications can also be built utilizing FSS, such as private search~\cite{huang2023multi, dauterman2020dory}, boolean queries~\cite{li2024nemo}, SQL~\cite{wang2017splinter}, oblivious RAM-related workloads~\cite{vadapalli2023duoram,doerner2017scaling,ji2023multi}.

Despite their strong online efficiency, 
FSS-based protocols incur substantial offline overheads that hinder practical deployment. 
For example, processing a single query using FSS-based secure inference~\cite{gupta2023sigma} on BERT-Base requires generating up to 16.84GB of FSS keys, and more than 200GB for LLaMA2-7B~\cite{gupta2023sigma,jawalkar2024orca}, which are distributed over the network (from a trusted third party). This incurs significant offline overheads that sometimes cannot be entirely hidden~\cite{huang2026beyond}, as well as storage overheads and I/O pressure (\autoref{sec:pattern}).
A single PIR query over a database of size $2^{20}$ requires over one million key expansion operations (\autoref{sec:pattern}), each involving multiple pseudorandom generator (PRG) operations, incurring significant online computation overheads~\cite{vadapalli2023duoram,doerner2017scaling,corrigan2015riposte,ji2023multi,lam2023gpu}. 

Our key insight is that, across diverse FSS-based workloads, the distributed point function (DPF) forms the dominant computational core, which can be further expanded to the distributed comparison function (DCF)~\cite{storrier2023grotto}. In secure inference, DPF is used inside most nonlinear operations, accounting for over 90\% of the non-linear computation and dominating offline key generation. In PIR, the DPF evaluation is performed across the entire database domain and becomes the main computational bottleneck. 

We built a lightweight distributed function accelerator (DFA) for DPF and co-designed the surrounding algorithm/system to accelerate multiple workloads that utilize FSS, including both latency-sensitive private inference and throughput-oriented PIR workloads.
At the hardware level, we design a (partially) programmable accelerator centered around a fixed-function DPF core that can accelerate several FSS-based workloads. At the algorithm level, we refactor the FSS protocols to decouple DPF generation and evaluation, enabling independent execution and efficient parallel generation. At the system level, we introduce an end-to-end execution model that eliminates unnecessary key storage requirements and data movement across CPU, GPU, and memory hierarchies.
To provide benefit across diverse threat models, our hardware provides two operating modes --- untrusted acceleration, which still provides some acceleration when the hardware is untrusted, and trusted acceleration, which provides even more acceleration when the hardware is equipped with additional secure computing capabilities and can be trusted for FSS key generation. 
Together, these techniques transform FSS from a theoretically appealing but system-heavy primitive into a practical building block for real-world privacy-preserving systems.

\change{
DFA is not intended to be novel as a standalone AES datapath or VPU. Its key insight is to organize simple hardware blocks around the FSS dataflow: fixed-function DPF support accelerates AES-heavy generation/evaluation, while the surrounding system changes execution from offline key generation, storage, transfer, and reread to GPU-side generate-and-consume. This organization reduces both computation for DPF-heavy workloads such as PIR and communication/ storage/ data movement for private inference.
}
This paper makes the following contributions:
\begin{itemize}[leftmargin=*]
    \item \textbf{Accelerator Hardware Design.}
    We design a DPF-centric accelerator that combines a high-throughput fixed-function DPF core with lightweight programmable protocol-specific processing. The design targets the dominant component in modern FSS workloads while preserving enough flexibility for higher-level FSS protocols.

    \item \textbf{Protocol Co-Design.}
    We refactor FSS protocols to decouple DPF generation from higher-level logic, and apply algorithm level optimization to reduce data storage and movements.

    \item \textbf{End-to-End System Integration.}
    We integrate the accelerator into multi-node FSS systems and support multiple deployment modes, including untrusted and trusted configurations, to address different security and system constraints while reducing communication, storage, and data movement overheads.

    \item \textbf{Evaluation.}
    We evaluate the proposed design using RTL implementations and system-level simulations, demonstrating substantial improvements in throughput, communication volume, and storage footprint across representative secure inference and PIR workloads. We achieve more than 10$\times$ offline communication reduction, storage reduction, end-to-end latency reduction, and around 5$\times$ energy savings for private inference; We achieve more than 5$\times$ PIR throughput and energy improvement compared to the previous GPU solution.
\end{itemize}

\section{Background}
\label{sec:bg}

\subsection{Function Secret Sharing (FSS) and Distributed Point Function (DPF)}
\label{sec:bg_dpf}

Function secret sharing (FSS) allows a function $f(\cdot)$ to be split into multiple function shares $(f_0(\cdot), f_1(\cdot))$, such that each party locally evaluates its share on an input, and the results combine to the output: with $y_0 = f_0(x), y_1 = f_1(x)$, we have $y = y_0 + y_1 = f(x)$. Each function share individually reveals no information about the original function, enabling secure distributed evaluation.

A commonly used special case of FSS is the distributed point function (DPF), which represents a point function (the $\bullet$ in the superscript means that this is a point function):
\[
f^{\bullet}_{\alpha,\beta}(x) =
\begin{cases}
\beta, & \text{if } x = \alpha \\
0, & \text{otherwise}
\end{cases}
\]
for a target index $\alpha$ and value $\beta$. DPF is widely used as a building block for secure selection and lookup operations.
%
Another main building block is the distributed comparison function (DCF), $f^{>}_{\alpha,\beta}(x)$, which outputs $\beta$ if $x>\alpha$, or 0 otherwise. Recent work~\cite{storrier2023grotto} showed that DCF can be efficiently constructed with DPF.
DPF and DCF can be used to build many other FSS functions~\cite{gupta2023sigma}.

This paper focuses on two-party FSS for simplicity, but the idea can be generalized to more parties.
Two-party FSS (including DPF, DCF, and other functions) works as follows. An offline key generation algorithm $\mathsf{Gen}(\alpha, \beta)$ produces two secret keys $(k_0, k_1)$. These keys are distributed to the two parties, and each key implicitly defines a function share via an evaluation algorithm, i.e., $f_i(x) = \mathsf{Eval}(k_i, x)$. The keys are generated such that for all $x$, the outputs satisfy $\mathsf{Eval}(k_0, x) + \mathsf{Eval}(k_1, x) = f(x)$, while each individual key reveals no information about $f$.
When the $\mathsf{Gen}$ and $\mathsf{Eval}$ functions are for DPF or DCF, we put $\bullet$ or $>$ in the superscript (i.e., $\mathsf{Gen}^\bullet$ and $\mathsf{Eval}^\bullet$ for DPF, and $\mathsf{Gen}^>$ and $\mathsf{Eval}^>$ for DCF). 

%

FSS protocols incur significant computation, communication, and storage overheads~\cite{huang2026beyond}.
Generating ($\mathsf{Gen}$) and evaluating ($\mathsf{Eval}$) the keys require repeated use of pseudorandom generators (PRGs), which internally use AES.
As FSS keys ($k_0, k_1$) are large and cannot be reused, in certain workloads, distributing them to parties over the network before each use incurs significant communication and storage overheads (on the receiving side)~\cite{huang2026beyond}.
%
In this paper, we focus on the DPF implementation established by Boyle et al.~\cite{boyle2016function}, a DCF implementation based on DPF~\cite{storrier2023grotto}, and other FSS function implementations that use DPF/DCF at its core~\cite{gupta2023sigma, jawalkar2024orca}. \change{For the two tasks we focus on in this paper, the FSS keys are one-time and cannot be reused across queries.}




\subsection{FSS-based Private Inference}\label{sec:bg_mpc}

Secure multiparty computation (MPC)~\cite{yao1982protocols} allows multiple parties to jointly compute a function on their inputs while keeping those inputs private. 
Recently, the use of MPC to privately run ML inference without revealing the input data or model weights is gaining attention~\cite{wagh2020falcon,kumar2020cryptflow,rathee2020cryptflow2,crypten2020,tan2021cryptgpu}.
In a common two-party MPC setting, a sensitive value $x$ is split into two random secret shares held by two non-colluding parties, $P_0$ and $P_1$, such that $[x]_0 + [x]_1 \equiv x$, with each share not revealing any information of the original secret. Additions can be computed without communication (e.g., to compute $z = x + y$, each party locally computes $[z]_i = [x]_i + [y]_i$), while multiplications and nonlinear operations require additional communication. These communication-heavy operations, especially complex non-linear operations (e.g., RELU, Softmax), become the dominant bottleneck when running ML inference with MPC.

Before FSS-based approaches, MPC-based secure inference mostly realized non-linear functions through arithmetic-to-binary (A2B) share conversion and bit-level operations~\cite{wagh2020falcon,mohassel2018aby3,demmler2015aby,wagh2019securenn,tan2021cryptgpu,crypten2020}, which required multiple rounds of communications. FSS-based approaches, which are relatively new, implement non-linear layers with FSS, using several comparisons (DCF) and lookup tables (LUTs; DPF) internally~\cite{gupta2023sigma, jawalkar2024orca, ryffel2020ariann} (\autoref{sec:goal}).
FSS-based private inference requires much less communication rounds than A2B-based.
Unfortunately, as discussed in \autoref{sec:bg_dpf}, using FSS has a hidden cost of having to generate, distribute, and store large FSS keys.
Prior works~\cite{jawalkar2024orca, gupta2023sigma} neglected these costs, assuming that they can be done offline and their overheads do not matter. However, a recent work~\cite{huang2026beyond} showed that the cost cannot be entirely hidden unless the query processing rate is very low, and these costs can dominate the performance and energy consumption.
%
In addition, online phase becomes more compute-intensive due to the key evaluation.

\subsection{Private Information Retrieval}\label{sec:pir}

Private information retrieval (PIR) allows a client to retrieve a record from a database without revealing the queried index to the server. 
Two common setups are single-server PIR and multi-server PIR.
Single-server PIR is often based on homomorphic encryption (HE) and incurs a higher computational cost~\cite{henzinger2023one}. Multi-server PIR~\cite{chor1998private,gilboa2014distributed} can be generally faster than single-server PIR, and this paper focuses on two-server PIR due to its relatively higher throughput, especially focusing on a popular DPF-based approach.

A typical DPF-based PIR works as follows. 
Given a query index $i \in [N]$, the client encodes it as a point function $f^{\bullet}_{i,1}(x)$, which is 1 only at $x=i$, and 0 otherwise. The client then generates two keys $(k_0, k_1) \leftarrow \mathsf{Gen}^{\bullet}(i,1)$ and sends the keys $k_j$ to the server $P_j$ for $j \in \{0,1\}$.
Each server locally evaluates the function over all possible indices, i.e., computes 
$\mathsf{Eval}^{\bullet}(k_j, x)$
for all $x \in [N]$. Given a database $\mathbf{D} = (D[1], \dots, D[N])$, each server then computes an inner product between its function share and the database:
\begin{equation}
    y_j = \sum_{x=1}^{N} \mathsf{Eval}^{\bullet}(k_j, x) \cdot D[x] \textrm{ for }j=0,1.
    \label{eq:pir}
\end{equation}
The client finally reconstructs the desired record by combining the responses ($D[i] = y_0 + y_1$). 

DPF-based PIR does not incur huge communication cost, because the key size is relatively small ($O(\log(N))$ for a database of size $N$).
%
However, running $\mathsf{Eval}^{\bullet}(k_j, x)$ over the entire $1 \le x \le N$ incurs a non-negligible compute overhead
of $O(N)$ PRG evaluations (\autoref{sec:pattern}) plus performing the inner product with the entire database.
Although cheaper compared to single-server PIR (HE-based), this is still compute-heavy even for modern GPUs~\cite{lam2023gpu}. 


\textbf{Scope.}
Throughout the rest of this paper, when we refer to \emph{private inference} and \emph{PIR}, we specifically mean two-party FSS-based private inference and two-server DPF-based PIR, respectively.



\section{System Overview}
\subsection{Threat Model}

FSS protocols are commonly designed for a semi-honest (honest-but-curious) adversary model~\cite{mohassel2018aby3,ryffel2020ariann,gupta2023sigma,jawalkar2024orca,mohassel2017secureml,wagh2019securenn,watson2022piranha,lam2023gpu,ji2023multi}, which assumes that all parties faithfully follow the protocol but may attempt to infer each other's secret. 
We adopt the same semi-honest adversary assumption.
We assume that the servers (multiple parties running private inference or PIR) do not collude, and the client does not collude with either server, which is a standard assumption in MPC/PIR literature~\cite{demmler2015aby,rathee2020cryptflow2,mohassel2017secureml,lam2023gpu,gupta2023sigma,jawalkar2024orca,dauterman2020dory}.
In the classical private inference workload without our DFA, we assume there exists a trusted third party that generates FSS keys and distributes them to other parties through the network.
%
Several other works in MPC~\cite{gupta2023sigma, crypten2020, liu2025depth, jawalkar2024orca} assumed a similar trusted third party.



We evaluate our proposed hardware under two threat models. The \emph{untrusted mode} supports the threat model of the traditional FSS protocols. No trust is placed on hardware and the security/privacy fully comes from the cryptographic protocol designs. For example, a party with physical access to the hardware is assumed to be able to access, or even tamper with, the data inside the proposed hardware.
%
\emph{Trusted mode} assumes that the proposed hardware provides data confidentiality and integrity, i.e., even the party physically holding the device cannot inspect (even through various side channels) or manipulate the internal state, similar to what is assumed for other trusted hardware such as trusted platform module (TPM)~\cite{tpm2spec} or trusted execution environment (TEE)~\cite{costan2016intel,arm_trustzone,amd_sev}.
Previous studies showed that having a small trusted hardware module can accelerate MPC~\cite{zhou2022ppmlac, huang2022efficient}. The trusted mode explores what additional benefits can be achieved for FSS under similar assumptions when a small secure hardware module can be trusted. 
\change{In the trusted mode, DFA is used as a third party for FSS key generation. Instead of relying on an external trusted dealer, each party uses a local trusted DFA to generate the same distribution of party-local keys. DFA keeps PRG seeds, counters, and intermediate key-generation state hidden from the host. The threat model for the on-line phase does not change from the underlying FSS/MPC protocol.}






\subsection{System Overview}
\begin{figure}
    \centering
    \includegraphics[width=\linewidth]{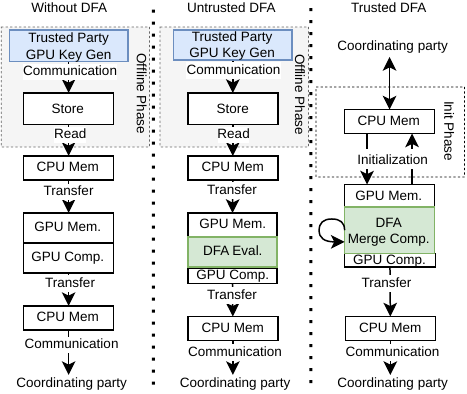}
    \caption{FSS acceleration using pure GPU solution without DFA, using DFA but untrusted, and using fully trusted DFA.}
    \label{fig:schemes}
\end{figure}
Our work focuses on two main applications: (1) FSS-based private inference~\cite{gupta2023sigma, jawalkar2024orca} and (2) DPF-based PIR~\cite{lam2023gpu}, which achieves the state-of-the-art in terms of online throughput in respective fields.
Our work can also be applicable to other DPF-based applications~\cite{huang2023multi, dauterman2020dory, li2024nemo, wang2017splinter, vadapalli2023duoram,doerner2017scaling,ji2023multi}.

We present an overview of our scheme compared to the baseline in~\autoref{fig:schemes}. 
Our main proposal introduces a new distributed function accelerator (DFA) that accelerates DPF (colored in green), and co-designs the algorithm/system around it to accelerate a larger class of FSS functions.
DFA is integrated as a small, dedicated accelerator on the GPU die and assists the GPU in handling DPF-related computation.
In the untrusted mode (\autoref{fig:schemes}, middle), DFA mainly serves as an engine for PRG that is required during the DPF evaluation phase, improving throughput and reducing energy consumption over a GPU-only implementation.
In the trusted mode (\autoref{fig:schemes}, right), DFA further reduces the communication overhead over the network by taking the role of the trusted third party in the baseline FSS protocol.
The trusted DFA enables local key generation, and minimizes the storage overhead and intermediate data transfers by generating and consuming keys on-the-fly, instead of storing them on local memory/SSD, effectively removing the offline phase.

For trusted mode, DFA is equipped with a secure hardware module that supports remote attestation and a key exchange protocol. Then, the DFA modules with a shared secret can generate FSS keys locally. In the trusted mode, each DFA generates FSS keys only for one party of the protocol while protecting the confidentiality and integrity of its operation as the `trusted third party'.

\section{Workload Analysis and Characterization}


\subsection{Core Insight: DPF as the Acceleration Target}\label{sec:insight}

Our core insight is that we can have a small, fixed function accelerator for DPF (DFA) and build the system around it to accelerate a larger class of FSS functions.
In this section, we briefly explain why accelerating DPF is important and crucial.

\subsubsection{Private Inference}

The major bottleneck is its offline key generation phase, where large volumes of keys must be generated, distributed, and stored in each party. For example, in private inference of BERT-base (with 128 tokens), prior work~\cite{gupta2023sigma} reports \textbf{16.84GB of FSS keys} to be generated and distributed. 
A recent prior work~\cite{huang2026beyond} showed that such a large offline cost may not always be hidden and may limit the inference latency/throughput.
Moreover, while the latency cost may be hidden if it can be done entirely offline, the energy or monetary cost cannot be hidden~~\cite{huang2026beyond}.



Our further analysis revealed that of the 16.84GB FSS keys, \textbf{over 90\% are DPF keys}, and the trend is similar for all other models in our experiments.
This is because the main overhead of MPC-based private inference lies in its non-linear operators (e.g., ReLU, Softmax, LayerNorm)~\cite{liu2025depth}, and DPF/DCF is heavily used to implement them with FSS.
MPC cannot natively support functions used inside these operators, like comparison, inverse, (inverse) square root, or exponential.
Hence, these functions are implemented by \emph{partitioning} the input range into several intervals and \emph{approximating} each interval's function behavior with low-degree polynomials or lookup tables~\cite{gupta2023sigma, pang2024bolt, ma2023secretflow}.
As DPF/DCF is used to obliviously determine in which interval the input lies, and also for comparison/LUT, they consist of the majority of the overheads.

\subsubsection{Private Information Retrieval}

As discussed in \autoref{sec:pir}, multi-server PIR also relies heavily on DPF.
%
The PIR computation pattern is rather simple: for each entry, a DPF evaluation result is followed by a multiplication (\autoref{eq:pir}).
In prior work~\cite{lam2023gpu} and in our experiments with NVIDIA A100, DPF evaluations achieve only around \textbf{20-50GBps throughput} (see \autoref{sec:eval} for setup details), which is significantly lower than multiplication throughput of modern GPUs, and becomes the main bottleneck (an NVIDIA A100 can sustain on the order of $10^{13}$ 32-bit integer operations, and $10^{11}$--$10^{12}$ 128-bit integer operations per second). 
Memory bandwidth is not the main bottleneck when the queries are batched, as one only needs to read the database once and perform multiple DPF/multiplications, making DPF the main bottleneck.
%



\textbf{Implications.}
As discussed, DPF is a major bottleneck for both private inference and PIR, making it a natural target for acceleration.
However, naively accelerating DPF does not automatically bring end-to-end speedup.
The detailed characteristics differ between the two workloads: in private inference, DPF calls are interspersed with other computations, while PIR calls DPF for a large number of evaluations to cover the entire database.
We have to carefully design the hardware to support both use cases and also co-design the surrounding system and algorithm, as we will discuss in \autoref{sec:design}.







\begin{figure*}[ht]
    \centering
    \includegraphics[width=1\linewidth]{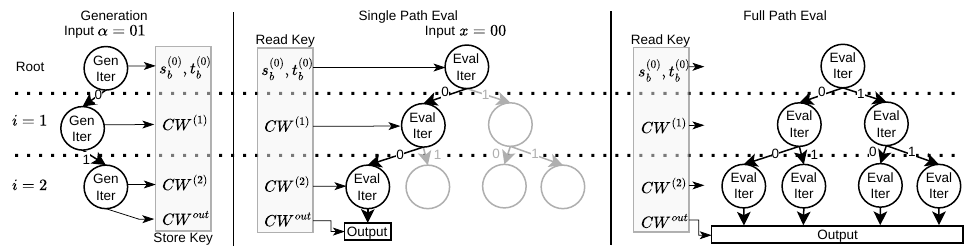}
    \caption{How the computation traverse through the binary tree of DPF during the generation and evaluation of different workloads, with tree depth being 2. $\alpha$ controls the path during the generation and $x$ controls the path during the single path evaluation. $b\in\{0,1\}$ represent the computing party index.}
    \label{fig:fss}
\end{figure*}

\begin{algorithm}[t]
\setstretch{0.78}
\caption{Simplified DPF Key Generation ($\mathsf{Gen}^\bullet_{\alpha, \beta}$)}
\begin{algorithmic}[1]\label{alg:gen}
\STATE \textbf{Input:} target index $\alpha \in \{0,1\}^n$, payload $\beta \in \mathbb{Z}_N$
\STATE \textbf{Output:} two DPF keys $k_0, k_1$

\STATE Sample random seeds $s_0^{(0)}, s_1^{(0)}\in \{0,1\}^{\lambda-1}$, $t_0^{(0)}, t_1^{(0)} \in \{0,1\}$

\FOR{$i=1$ to $n$}
    \STATE 
    $(s_b^L,t_b^L,s_b^R,t_b^R) \leftarrow \mathsf{PRG}(s_b^{(i-1)}) \quad \text{for } b\in\{0,1\}$
    \STATE $\mathsf{CW}^{(i)}\leftarrow \textsf{Compute\_CW}(\{s_b^L,t_b^L,s_b^R,t_b^R\}_{b\in\{0,1\}},t_1^{(i-1)},\alpha[i])$
    \STATE $s_0^{(i)}, t_0^{(i)}, s_1^{(i)}, t_1^{(i)}\leftarrow \textsf{Compute\_seed}({CW}^{(i)})$
\ENDFOR

\STATE $\mathsf{CW}^{\mathrm{out}} \leftarrow \textsf{Final\_layer}(s_0^{n},t_0^{n}, s_1^{n}, t_1^{n},\beta)$.
\STATE $k_0 \gets (s_0^{(0)}, t_0^{(0)}, \{\mathsf{CW}^{(i)}\}_{i=1}^n, \mathsf{CW}^{\mathrm{out}})$
\STATE $k_1 \gets (s_1^{(0)}, t_1^{(0)}, \{\mathsf{CW}^{(i)}\}_{i=1}^n, \mathsf{CW}^{\mathrm{out}})$
\STATE \textbf{return} $k_0, k_1$
\end{algorithmic}
\end{algorithm}

\begin{algorithm}[t] 
\setstretch{0.78}
\caption{Simplified DPF Evaluation ($\mathsf{Eval}^\bullet_{\alpha, \beta}$)}\label{alg:eval}
\begin{algorithmic}[1]
\STATE \textbf{Input:} one DPF key $k_b$, query index $x \in \{0,1\}^n$
\STATE \textbf{Output:} one share $y_b \in \mathbb{Z}_N$

\STATE Parse $k_b$ as root seed $s^{(0)}$, control bit $t^{(0)}$, level correction words $\{\mathsf{CW}^{(i)}\}$, and final correction $\mathsf{CW}^{\mathrm{out}}$

\FOR{$i=1$ to $n$}
    \STATE 
    $(\hat s^L,\hat t^L,\hat s^R,\hat t^R) \leftarrow \mathsf{PRG}(s^{(i-1)})$
    \STATE $(s^L,t^L,s^R,t^R)\leftarrow \textsf{Compute\_CW}(\hat s^L,\hat t^L,\hat s^R,\hat t^R, \mathsf{CW}^{(i)},t^{(i-1)})$
    \STATE $(s^{(i)},t^{(i)}) \gets (s^L,t^L)$ if $x[i]=0$, else $(s^R,t^R)$
\ENDFOR

\STATE $y_b \leftarrow \textsf{Final\_output}(\mathsf{CW}^{\mathrm{out}},s^{n}, t^{n},b)$
\STATE \textbf{return} $y_b$
\end{algorithmic}
\end{algorithm}
\subsection{Compute Pattern of DPF}

Given a target index $\alpha \in \{0,1\}^n$, the goal is to generate two keys $k_0, k_1$ such that, for any input $x$, two servers can evaluate shares of a point function that outputs $\beta$ if $x=\alpha$ and $0$ otherwise. We show an illustration in~\autoref{fig:fss}, and show the pseudocode in \autoref{alg:gen} (generation) and~\autoref{alg:eval} (evaluation). As can be seen in step 5 of both algorithms, the PRG ``expands'' the seed to roughly double its length, which is why we call PRG operations ``key expansion''.

\textbf{High-level idea.}
DPF can be viewed as a binary tree. The DPF keys define the tree and are generated during \textbf{generation} phase. As shown in the generation section in~\autoref{fig:fss} (left), $\alpha$ defines a unique path from the root to the leaf by selecting the path it traverses through the binary tree. A pair of root seeds $s_{b}^{(0)},t_{b}^{(0)}$ for $b=0,1$ is randomly sampled (as in \autoref{alg:gen} step 3). In each generation iteration $i$, the seeds are expanded by PRG (step 5), and the results are selected by the $i$th digit of $\alpha$ to compute a correction word $\mathsf{CW}^{(i)}$ (step 6), which is used to determine the seeds $s_{b}^{(i)},t_{b}^{(i)}$ for the next iteration (step 7). The keys are all correction words (each $O(\lambda)$ length) and the root seeds combined (step 10, 11).



During \textbf{evaluation}, $\mathsf{Eval}^\bullet(k_b, x)$ takes one party's key $k_b$ and an input index $x \in \{0,1\}^n$. $x$ indicate the paths that evaluation traverses through in the binary tree (signle-path eval; \autoref{fig:fss}, middle), and the key is used to compute whether $x$ follows the path defined by $\alpha$. The key is parsed to recover all $\mathsf{CW}$s and root seeds (step 3 of~\autoref{alg:eval}), and in each iteration $i$, the current seed gets expanded (step 5), is processed with $\mathsf{CW}^{(i)}$ (step 7), and then the next seeds are selected by the $i$th bit of $x$ (step 10). Eventually, the output will be shares of $\beta$ if and only if the path is identical to the path in generation, which is $x=\alpha$. In the full-path eval, no $x$ is given, and for each node both 0 and 1 are selected in step 7, and both outputs are computed in the next iteration (full-path eval; \autoref{fig:fss}, right).


\textbf{Role of the PRG.}
The main computation in DPF is repeated seed expansion. 
At each tree level, a seed $s$ is expanded into two pseudorandom child seeds in step 5 of both~\autoref{alg:gen} and ~\autoref{alg:eval}.
In practice, the PRG is built from AES-based pseudorandom expansion, so DPF generation and evaluation are dominated by repeated AES calls, followed by lightweight XOR/select logic (discussed in~\autoref{sec:results}).


\begin{figure}[t]
    \centering
    \includegraphics[width=1\linewidth]{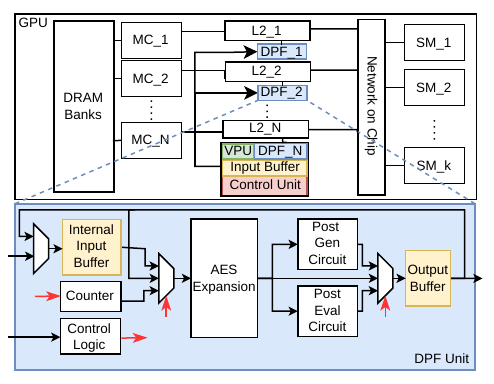}
    \caption{The DFA and inside GPU (top) and DPF unit (bot).}
    \label{fig:dfa}
\end{figure}

\subsection{Workload Pattern Analysis}\label{sec:pattern}

Different use cases utilize DPF with different patterns. There are mainly two ways, single-path evaluations and full-path evaluations.

Private inference mostly perform \textbf{single-path evaluation}, which only evaluates one root-to-leaf path along the DPF tree (\autoref{fig:fss}, middle). This is because private inference uses DPF/DCF to figure out on which interval a particular input lies or to perform something like LUT/comparison for a single input.
Single-path evaluation have relatively lower online computation, requiring $n$ PRG key expansions along one path of an $n$-bit input.
The relative offline key generation overhead becomes larger, because the key must be discarded after one path evaluation.
As a result, the major bottleneck of private inference is in generating, distributing, and storing keys during the offline phase, as discussed in \autoref{sec:bg_mpc}.
Each key size scales with $n$ and with the security parameter $\lambda$ ($O(\lambda n)$).

PIR workloads perform \textbf{full-path evaluation}, which evaluates all paths in the DPF tree (\autoref{fig:fss}, right), because it must evaluate the point function for all possible indices (\autoref{eq:pir}).
The online computational complexity is much higher ($O(2^n)$, where $2^n$ is the database entry size) compared to single-path evaluation, and the intermediate state during the online evaluation is also large, pressuring memory resources.
The offline key generation is relatively less of an issue, as one key is reused for $O(2^n)$ evaluations (for single-path inference, it is discarded after one evaluation).
While full-path evaluation is sometimes used in private inference as well, private inference uses dominantly single-path evaluation.

\section{Architecture and System Design}\label{sec:design}
\subsection{Design Insights}\label{sec:goal}

Our goal is to design a hardware accelerator (distributed function accelerator, or DFA) and its surrounding system/algorithm to support two main FSS-based workloads, secure inference and PIR, while making sure the design is flexible enough for future workloads.

%

\textbf{Key insight.}
As discussed in \autoref{sec:insight}, both workloads heavily rely on DPF key generation and evaluation at its core, and accelerating DPF would like to accelerate both workloads of interest.
However, as discussed in \autoref{sec:pattern}, the two workloads differ in the computational pattern (single-path vs. full-path). Also, the hardware must be flexible to support other FSS functions, which relies on DPF/DCF at its core, but still have other computations.


We therefore adopt a hybrid architecture: A \textbf{fixed-function DPF engine} that accelerates the DPF operations (dominated by AES-based PRG), which is the main bottleneck, and a \textbf{small programmable unit} (vector processing unit, or VPU) that handles protocol-specific pre/post-processing logic and higher-level orchestration.
%
The fixed-function DPF engine provides high throughput by accelerating the dominant DPF computations, while the programmable VPU preserves the ability to support general, evolving FSS protocols. 
As long as the protocol is built around DPF/DCF, like modern private inference and PIR workloads, our DFA hardware can adopt to it by reprogramming the VPU, without having to redesign and redeploy a totally different hardware. \change{DFA is therefore most effective for DPF/DCF-heavy protocols, and other workloads dominated by non-DPF logic would rely more on the VPU/GPU path and benefit less from the DPF engine.}


We further observe that tight integration with the GPU is critical for both workloads. In the single-path workload, non-linear function evaluations (in ML inference) often have linear layers in between, which introduce substantial volumes of matrix and vector operations that are most efficiently handled on the GPU. In the full-path workload, DPF expansion is immediately followed by large matrix-vector multiplications, again requiring GPU for high throughput. In both cases, the computation involves frequent, and sometimes high-volume, data exchanges between our DFA and the GPU. To minimize this overhead and fully leverage data locality, we design DFA as a small on-die acceleration unit within the GPU architecture. These observations motivate our design.

















\begin{figure}[ht]
    \centering
    \includegraphics[width=\linewidth]{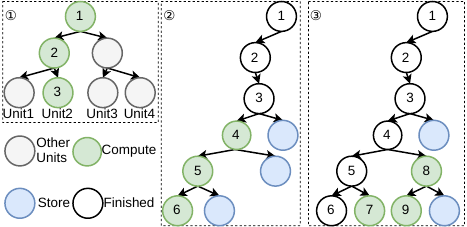}
    \caption{The computation pattern of 4 DPF units evaluating the full DPF tree of depth 5. 
    Each unit will compute one subtree below the four nodes in level 2 (\textcircled{1}).
    \textcircled{2} and \textcircled{3} shows the computation of unit 2, with numbers indicating the traversal order. The top-level repeated computation is negligible compared to the total computation in real applications.}
    \label{fig:dpf_compute}
\end{figure}
\subsection{Hardware Design}

DFA is a hybrid centralized--distributed organization: (1) \emph{distributed DPF units} provide high-throughput fixed-function acceleration near memory, and (2) \emph{centralized control unit and VPU} handle trusted initialization coordination and protocol-specific logic (\autoref{fig:dfa}). This separation enables efficient execution of the dominant DPF computation while maintaining flexibility for evolving FSS protocols.
DPF units are distributed and co-located with L2 banks, while the control unit and VPU are placed next to one designated L2 partition. SMs interact with the accelerator through the standard memory interface, allowing seamless integration into the GPU. 

\textbf{DPF Unit.}
DPF units accelerates PRG key expansion for seed generation and tree traversal.  
Each unit is a fixed-function pipeline centered around a deeply pipelined AES-based PRG engine. Control logic orchestrates per-level expansion and selection, while lightweight post-processing (128-bit ADD/XOR) applies correction words and completes circuit evaluation. Intermediate states are buffered in small on-chip SRAM to sustain streaming execution.  
The pipeline has a depth of 20 stages and sustains one Gen/Eval iteration per cycle after warmup.

Each DPF unit would process a single path at a time for single-path workloads independently, immediately forwarding intermediate states to the next level. For full-path workloads, the unit follows a depth-first traversal as shown in~\autoref{fig:dpf_compute}. When two child results are generated, one is stored and later revisited (blue nodes in \autoref{fig:dpf_compute}) while the other is immediately processed (green nodes in \autoref{fig:dpf_compute}). This reduces total memory usage from $O(2^n)$ to $O(nm)$ compared to breadth-first traversal, where $m$ is the number of DPF units. The stored nodes are revisited after completing its subtree. In our implementation, we use 16 DPF units with 64 AES-128 engines.

\textbf{Control Unit.}
The control unit manages system initialization and trusted operations.  
It performs attestation, secure key provisioning, and task setup. We modeled this unit after an ARM SecurCore SC300~\cite{st33tpm12spi}, a security-hardened microcontroller based on Cortex-M3~\cite{arm_cortex_m3}, providing sufficient programmability and cryptographic support.
At initialization, the two DFAs (one per party) exchange parameters, mutually attest, and establish a shared secret key. This key remains inaccessible to the host system and is used internally to synchronize pseudorandom generation.


\textbf{Programmable Vector Processing Unit (VPU).}
The VPU handles non-DPF components of FSS protocols, including protocol-specific transformations and masking.  
Compared to the DPF unit, the VPU provides lower throughput but still sufficient compute capacity, as non-DPF operations are not the dominant cost (shown in~\autoref{sec:results}).  
\emph{Interaction.} Separation between the DPF unit and the VPU is enabled by algorithm-level decoupling of DPF and non-DPF computations (discussed in~\autoref{sec:alg}), allowing both components to operate in parallel without introducing bottlenecks.


\textbf{Memories and Communication Interface.}
%
\change{Each memory partition contains an L2 bank and a local DPF engine with its output buffer behind the same NoC-facing interface. The DPF engine is not part of the L2 cache and does not use the L2 tag/data arrays.} 
\change{Normal memory requests are forwarded to the L2 bank, while requests to a reserved DFA address range are forwarded to DFA control registers or DFA-local output buffers. 
During execution, the runtime writes input for DPF units in the input buffer, and they are then forwarded to the DPF engines with dedicated links. Read requests to the output buffers from SMs arrive through the existing NoC link. SMs consume generated outputs by issuing loads to the DFA output address range, which are served from the DFA output buffers and returned through the same shared NoC-facing interface. Thus, DFA does not add a separate NoC port, and it also does not require normal L2 accesses to pass through the DPF engine.
}

To sustain throughput, the system employs double buffering for both input/output, enabling overlap between data movement and computation. The internal buffers within each DPF unit maintain a transient state during execution.
Single-path workloads require less than 1KB of internal buffering (full-path workloads do not require trusted mode). As a result, the trusted computing base remains compact despite the larger buffers used for full-path evaluation.

\textbf{Dataflow, Storage, and Backpressure.}
\change{Each DPF unit writes generated outputs to its local output buffer. 
In trusted generate-and-consume mode, DPF keys and intermediate states are consumed inside DFA and discarded. In that sense, the DPF keys are not exposed to external modules such as CPU memory, GPU memory, L2, or global memory. Temporary storage is determined by pipeline depth, in-flight scalar DPF instances, and buffering for SM--DFA access latency, not by model dimension; larger models mainly increase the number of scalar DPF invocations. If the VPU or SM consumer falls behind, the corresponding DPF pipeline stalls once its output buffer fills. During generation, DPF generation is decoupled from VPU preprocessing, so a slow VPU limits generation throughput but does not block DPF engine throughput.}
\change{Additional SM--L2 NoC traffic is proportional to the DFA-generated outputs consumed by SMs. For example, a 1M-entry PIR query reads about 16MB of DFA output ($\sim$134K 128B cache-line transfers), while BERT-base evaluation moves about 300MB ($\sim$2.3M transfers). This traffic and contention are modeled in Accel-Sim through the shared NoC paths.}

\textbf{Design-space Comparison.}
\change{We compare the L2-adjacent DFA placement with several alternative DPF/AES acceleration points.}

\textit{\change{CPU AES or off-chip PRNG acceleration.}}
\change{This option can accelerate AES/PRNG expansion, but the generated DPF/FSS stream still has to be transferred to the GPU.}
\change{On an A100 PCIe Gen4 x16 system, one-way PCIe bandwidth is about 31.5GBps, while DFA provides roughly 290GBps of DPF/FSS bandwidth to support our use cases.}
\change{Thus, CPU/off-chip generation would provide only about 9--11\% of throughput.}

\textit{\change{Single centralized DFA engine.}}
\change{If the same 64 AES engines were placed in one centralized engine, the generated stream would have to leave through a single attachment point.}
\change{With one memory-channel path of about 180GBps peak bandwidth, this would be capped to only about 62\% of the 290GBps performance.}

\textit{\change{GPU AES instructions inside SMs.}}
\change{This design requires ISA/programming-stack changes.}
\change{With one AES engine per 108 SMs, this requires 108 AES engines, 1.69$\times$ more compared to our design that places DPF units next to each L2 (64 AES engies).
Moreover, using SMs for DPF key generation makes it difficult to support the trusted mode because it requires the entire SM to be included in the trusted boundary.}

\textit{\change{DFA next to each SM.}}
\change{A per-SM DFA can be designed to run decoupled from the main SM, the tradeoff is replication and sharing granularity.}
\change{If each per-SM DFA is local-only, DPF capacity is statically partitioned across SMs. When an SM is occupied and cannot handle the postprocessing after DFA generation, the DPF engines for that SM cannot be used.
}
\change{The per-SM DPF unit also leads to larger area overhead. For our configuration, adding 1 AES engine per SM requires 108 AES engines.
Moreover, reducing each local unit to one AES engine under-utilizes the DPF post-processing pipeline, because the post processing needs to wait for multiple rounds for all inputs from AES being ready. This requires additional intermediate-state registers; our estimate is that this adds about 5\% area to each DPF unit. The design is overall 1.74$\times$ in area compared to our design.}

\change{Overall, compared to our design, CPU/off-chip generation is limited by CPU--GPU transfer, a centralized engine is limited by output bandwidth and routing, SM-based AES enlarges the trusted boundary, and per-SM DFA requires more hardware.}

\subsection{Algorithm Design}\label{sec:alg}

\textbf{Decoupling DPF Generation.}\label{sec:decoup}
Existing FSS-based private inference tightly couple DPF key generation with protocol-specific preprocessing. In specific non-linear function protocols, like DReLU~\cite{gupta2023sigma}, the inputs to $\textsf{Gen}^\bullet$ are derived from intermediate values computed during higher-level logic (specifically, the target index $\alpha$ in ~\autoref{alg:gen}). In such cases, the fast DPF units will have to wait for the slow VPU pre-processing, which brings an extra bottleneck. 

We observe that, in FSS-based protocols for non-linear functions, the inputs to $\textsf{Gen}^\bullet$ are typically uniformly random $r \in \mathbb{Z}_N$, or obtained via simple arithmetic transformations of $r$ as $g(r)$. Most of these transformations are bijective over the ring and preserve uniformity, including only addition and XOR. This is fundamentally required by the DPF generation algorithm, whose security argument comes from the fact that its inputs are uniformly random. For the few cases where FSS generation involving non-bijective integer operations like multiplication, SOTA protocols~\cite{gupta2025shark, gupta2023sigma, jawalkar2024orca, wu2025epbnn} show that those functions do not use DPF/DCF.

We leverage this fact that protocol-level transformations are bijective and preserve uniformity to decouple DPF generation from protocol-specific pre-processing. Instead of deriving inputs from higher-level logic, we directly sample a uniformly random value $\tilde{r} \in \mathbb{Z}_N$ (e.g., using an PRG) and perform DPF key generation independently without waiting for pre-processing.
When the protocol requires the original value $r$, it can be recovered via the inverse transformation $g^{-1}(\tilde{r})$, preserving both correctness and distribution. A special case is right shift (truncation), which has no strict inverse, but we can keep the (computationally equivalent) uniform randomness by doing left shift and fill the lower bits with uniform random sampling. The proof of security follows directly from a standard simulation-based proof, as the values observable by the parties are computationally indistinguishable from uniform randomness.

To the best of our knowledge, typical FSS protocols do not leverage DPF keys for direct post-processing beyond the corresponding DPF evaluation. This is because the keys are carefully structured to encode only the information necessary for traversing the DPF tree, and are therefore usable only within the corresponding evaluation procedures. Our above transformation isolates DPF expansion as a standalone component from pre-processing, allowing it to be executed independently of protocol-specific pre-processing logic which runs on VPU. The generation of keys of the non-DPF portion can therefore be amortized during the whole generation period in parallel to the DPF portion, enabling sustained pipeline throughput. 



\subsection{System Integration}\label{sec:system}
\textbf{Merging Offline-Online.}
In prior FSS-based inference systems, the offline phase is typically handled by a trusted third party (TTP) that generates large volumes of FSS keys and distributes them to the evaluation parties. This design introduces three major inefficiencies. First, the generated keys incur substantial communication overhead during distribution. Second, the keys must be stored, leading to significant disk and memory footprint. Third, these precomputed keys must be transferred from CPU memory (or disk) to GPU memory for evaluation, resulting in considerable data movement overhead and potential bandwidth bottlenecks.

With our trusted secure hardware accelerator, the first issue is largely eliminated: key generation can be performed locally, and only the key material required by each party is exposed within the local trust boundary, removing the need for explicit key distribution. However, the latter two challenges remain. Because generation and evaluation are still decoupled, keys are produced in bulk without immediate consumption, requiring them to be buffered in memory or storage. This creates pressure on both CPU and GPU memory. 

We address these limitations by merging DPF key generation and evaluation into a single generate-and-consume execution flow, and the separation between offline and online phases is no longer necessary. Instead of first generating and storing full DPF keys and then transferring them during evaluation, the PRG key expansion used for evaluation is directly driven by the same PRG stream used during generation, and intermediate values are consumed immediately once generated. We show the details in~\autoref{alg:merged_dpf}

This design avoids moving the keys out of the compute engines or intermediate tree states whenever possible. All pseudorandom values are derived from a shared secret key and counter, including the target index $\alpha$ which we discussed in~\autoref{sec:decoup}.  $\mathsf{CW}^{(i)}$, the main part of the key, is immediately consumed after generation. Only the query index $x$, a shared key $K$ (which is established during the initialization phase and can be reused), and a counter state are required as inputs to drive the merged procedure.

The merged execution does not need explicit communication for key distribution from the TTP, and immediate consumption is provided by the evaluation logic as values are generated.
As a result, it reduces the key distribution communication and storage footprint, and extra data movement overhead between the generation and evaluation stages, as shown in~\autoref{fig:schemes}.






\subsection{Security}\label{sec:security}

\textbf{Untrusted vs. Trusted Operation.}
As illustrated in~\autoref{fig:schemes}, DFA supports both untrusted and trusted deployment modes. In the \emph{untrusted mode} (middle), the accelerator is used purely for performance; it executes DPF expansion on pseudorandom inputs, but all security guarantees remain the same as the baseline FSS protocol. 
In the \emph{trusted mode}, DFA 
acts as a trusted third party that generates FSS keys.
Remote attestation allows each party to verify that the other DFA module participating in an FSS protocol runs the correct code for the trusted party. Then, at run-time, DFA protects the confidentiality and the integrity of the key generation from the rest of a system. 
As long as the DFA-based key generation is correct, there is no change to the baseline FSS protocol.



\textbf{Lightweight Trusted Hardware Security.}
\change{
Lightweight trusted hardware has been widely deployed in security-critical systems. Examples include discrete security chips such as TPMs~\cite{pearson2003trusted}, Google Titan~\cite{googletitan}, and Apple T1, as well as earlier dedicated security devices such as smartcards~\cite{rankl2004smart} and IBM 4758~\cite{ibm4758}. Modern SoCs also commonly integrate dedicated security subsystems, including Synopsys tRoot~\cite{Synopsys}, Rambus RT-630~\cite{Rambus}, Apple Secure Enclave~\cite{AppleS}, and Qualcomm SPU~\cite{QSPU}. These designs typically provide secure boot, device identity, attestation, secure key storage, and hardware cryptographic engines, supporting a small trusted boundary with stronger isolation than a general-purpose CPU/GPU TEE.
}

\change{DFA, as a lightweight trusted hardware, is a narrower assumption than a full CPU/GPU TEE because the trusted boundary is limited to the small dedicated DFA, rather than a general-purpose processor with speculative execution, complex cache hierarchy, OS/runtime stack, or external memory system. Many leakage sources studied for high-performance TEEs are therefore outside the DFA trusted boundary, including page-fault/address-translation channels~\cite{xu2015controlled,shinde2015preventing}, shared-cache attacks~\cite{gotzfried2017cache,brasser2017software}, branch-prediction/transient-execution attacks~\cite{lee2017inferring,lee2017hacking,van2018foreshadow}, and DRAM-based attacks such as Rowhammer~\cite{kim2014flipping}. For DFA, timing and interconnect leakage are also limited since most intermediate values are pseudorandom.
However, small trusted hardware can still have vulnerabilities~\cite{moghimi2020tpm,han2018bad,butterworth2013bios}, and physical attacks such as probing, fault injection, power, or electromagnetic analysis require appropriate hardening~\cite{quadir2016survey,skorobogatov2017microprobing,skorobogatov2003optical}. These attacks and potential implementation bugs in the AES/PRG engine or control logic require standard hardware hardening and validation to prevent, similar to other lightweight trusted hardware and security processors~\cite{moghimi2020tpm,han2018bad}.}


\textbf{Protocol.} The protocol refactoring in~\autoref{sec:alg} (decoupled DPF generation and merged generate-and-consume execution) preserves the standard security assumptions. These transformations preserve functionality and output distribution: intermediate values and final outputs remain computationally indistinguishable from uniformly random. 
While we did not include a formal proof due to space limit, the proof follows standard simulation-based arguments in MPC/FSS and does not introduce additional leakage. \change{Untrusted DFA should be viewed mainly as an energy- and computation-throughput-oriented optimization for DPF execution; for private inference, it does not substantially reduce latency or communication because offline key generation, key distribution and movement remain as the bottleneck. The full cryptographic security proof will be included in the full version.}

\textbf{Security and Side-Channel Considerations.}
The accelerator is compact and the trusted compute base (the DFA) primarily executes regular, data-independent cryptographic operations (e.g., PRG and simple integer logic). Unlike other TEE-based systems that process plaintext (or directly derived) values within trusted hardware, our design operates mainly on pseudorandom values to derive FSS key material. This may reduce the exposure to data distribution-dependent leakage within the trusted computing base. A comprehensive analysis is still needed for future work.

\textbf{Trusted boundary and attestation.}
\change{DFA trusted boundary includes the device identity key, attestation logic, secure key storage, counters/nonces, AES/PRG engine, intermediate PRG/DPF state, and correction words before release. The host CPU, OS, driver, GPU SMs, L2/cache hierarchy, HBM/DRAM, network stack, and normal GPU memory are not trusted. During initialization, DFA uses measured boot and remote attestation to verify the firmware/configuration, then performs authenticated key exchange through the untrusted network to establish PRG seeds. The host cannot forge attestation or learn the seeds. Failed attestation leads to abort.}


\begin{algorithm}[!t]
\setstretch{0.78}
\caption{Merged DPF Generate-and-Consume Execution}
\label{alg:merged_dpf}
\begin{algorithmic}[1]
\STATE \textbf{Input:} query index $x \in \{0,1\}^n$, shared key $K$, counter $ctr$.
\STATE \textbf{Output:} output share $y_b \in \mathbb{Z}_N$
\STATE Sample target index: $\alpha \in \{0,1\}^n \leftarrow \mathsf{PRG}_K(ctr)$, 
\STATE Set Gen root states:  
\STATE Set Eval root states: $(\tilde s_0^{(0)}, \tilde t_0^{(0)},\tilde s_1^{(0)}, \tilde t_1^{(0)}) := (s_0^{(0)}, t_0^{(0)},s_1^{(0)}, t_1^{(0)})$
\FOR{$i=1$ to $n$}
    \STATE \textbf{(Gen Iteration Starts)} Expand generation states: $(s_b^L,t_b^L,s_b^R,t_b^R) \leftarrow \mathsf{PRG}_K(s_b^{(i-1)}) \quad \text{for } b\in\{0,1\}$
    \STATE $\mathsf{CW}^{(i)}\leftarrow \textsf{Compute\_CW}(\{s_b^L,t_b^L,s_b^R,t_b^R\}_{b\in\{0,1\}},t_1^{(i-1)},\alpha[i])$
    \STATE $s_0^{(i)}, t_0^{(i)}, s_1^{(i)}, t_1^{(i)}\leftarrow \textsf{Compute\_seed}({CW}^{(i)})$
    \STATE \textbf{(Eval Iteration Starts)} Expand the current evaluation state: $(\hat s^L,\hat t^L,\hat s^R,\hat t^R) \leftarrow \mathsf{PRG}_K(\tilde s^{(i-1)})$
    \STATE $(\tilde s^L,\tilde t^L,\tilde s^R,\tilde t^R)\leftarrow \textsf{Compute\_CW}(\hat s^L,\hat t^L,\hat s^R,\hat t^R, \mathsf{CW}^{(i)},t^{(i-1)})$
    \STATE $(\tilde s^{(i)},\tilde t^{(i)}) \gets (\tilde s^L,\tilde t^L)$ if $x[i]=0$, else $(\tilde s^R,\tilde t^R)$
\ENDFOR
\STATE $\mathsf{CW}^{\mathrm{out}} \leftarrow \textsf{Final\_layer}(s_0^{n},t_0^{n}, s_1^{n}, t_1^{n},\beta)$
\STATE $y_b \leftarrow \textsf{Final\_output}(\mathsf{CW}^{\mathrm{out}},\tilde s^{n}, \tilde t^{n},b)$
\STATE \textbf{return} $y_b$
\end{algorithmic}
\end{algorithm}


\begin{table*}[ht]
\centering
\caption{Single-path (inference) workload, with pure GPU solutions vs GPU + trusted DFA solutions with metrics measured per query. GPU energy numbers are the sum across the two computing parties and the trusted third party when present. Latency numbers are end-to-end including offline overhead. \change{GaC refers to the generate-and-consume optimization.}}
\vspace{-8pt}
\label{tab:single_path_th}
\small
\begin{tabular}{l l c c c c c}
\hline
\textbf{Model} & \textbf{Method} & \textbf{Offline Comm (GB)} & \textbf{CPU--GPU (GB)} & \textbf{LAN Latency (s)} & \textbf{WAN Latency (s)} & \textbf{GPU Energy (J)} \\
\hline
\multirow{4}{*}{ResNet-50}
& GPU & 9.9 & 10.2 & 10.9 & 162 & 71.0 \\
& GPU+Untrusted DFA & 9.9 & 10.2 & 10.6 & 162 & 11.8 \\
& \change{GPU+Trusted DFA w/o GaC} & \change{0.02} & \change{20.0} & \change{1.35} & \change{18.6} & \change{11.9} \\
& GPU+Trusted DFA & 0.02 & 0.34 & 0.73 & 18.0 & 11.9 \\
\hline
\multirow{4}{*}{BERT-Base}
& GPU & 17.1 & 19.3 & 18.8 & 296 & 121 \\
& GPU+Untrusted DFA & 17.1 & 19.3 & 18.4 & 296 & 18.1 \\
& \change{GPU+Trusted DFA w/o GaC} & \change{0.1} & \change{36.1} & \change{2.47} & \change{47.4} & \change{18.2} \\
& GPU+Trusted DFA & 0.1 & 2.55 & 1.42 & 46.4 & 18.2 \\
\hline
\multirow{4}{*}{GPT-Neo}
& GPU & 76.4 & 80.4 & 83.6 & 1140 & 587 \\
& GPU+Untrusted DFA & 76.4 & 80.4 & 82.1 & 1139 & 112 \\
& \change{GPU+Trusted DFA w/o GaC} & \change{0.24} & \change{156} & \change{10.7} & \change{106} & \change{114} \\
& GPU+Trusted DFA & 0.24 & 4.27 & 5.93 & 101 & 114 \\
\hline
\multirow{4}{*}{LLaMA2-7B}
& GPU & 256 & 268 & 279 & 3695 & 1879 \\
& GPU+Untrusted DFA & 256 & 268 & 273 & 3692 & 299 \\
& \change{GPU+Trusted DFA w/o GaC} & \change{0.81} & \change{523} & \change{34.0} & \change{209} & \change{303} \\
& GPU+Trusted DFA & 0.81 & 12.8 & 18.0 & 193 & 303 \\
\hline
\end{tabular}
\end{table*}

\section{Evaluation}\label{sec:eval}

\subsection{Evaluation Methodology}
We use RTL synthesis to obtain area, frequency, and power of the proposed hardware, cycle-level simulation to model system-level integration and data movement, and real-system runs to collect baseline execution time and GPU power for non-FSS components.
We report: \textbf{area} of the combined design (AES engine, control logic, SRAM, and VPU); peak DPF generation \textbf{throughput} from the synthesis results; estimated \textbf{energy} cost per query based on activity factors and measured/simulated power models.

\textbf{Hardware Cost.}
We wrote and synthesized a full RTL implementation of the DPF units, including
the DPF engine itself and all on-chip SRAM buffers. We use results from previous published implementations or reports for the VPU and the control core.
The AES engine is based on an open-source pipelined AES-128 implementation from OpenCores~\cite{opencores_aes128_pipeline} and is integrated as a fixed-function unit. Although more efficient implementations are available in academic work~\cite{dong201945nm}, we use the open-source version for RTL synthesis \change{for reproducible and conservative synthesis. A more optimized AES pipeline would likely further improve DFA performance and energy efficiency, but our main conclusion does not rely on highly optimized AES.} The control and preprocessing logic is synthesized to capture the overhead of protocol-specific operations and scheduling. SRAM buffers are sized to support the required throughput of the DPF pipeline.

All custom components are synthesized using a TSMC 28nm technology library and standard synthesis tools. Although the technology node is not state-of-the-art, it provides a consistent and reproducible baseline for the estimation of area, frequency, and power. We use DeepScaleTool~\cite{sarangi2021deepscaletool} to perform a tech node transformation to show the estimated metrics on newer technology.
For VPU, we leverage \textit{Ara}~\cite{cavalcante2019ara}, an open-source RISC-V VPU, with 16 lanes. Ara provides publicly available RTL and detailed reports on area, frequency, and performance. We use these reported numbers directly to model the overhead of deploying such a unit.

\textbf{System-Level Performance.}
To evaluate end-to-end performance, we built a system-level model that captures the interaction between DPF generation, VPU, and GPU. We use the privacy parameter $\lambda=127$ and the bit width of 64 for secret shares in all experiments.
For non-DPF FSS operations, we use the open-source Ara~\cite{cavalcante2019ara} simulator to obtain cycle-level estimates of execution latency. These cycle counts are combined with the reported operating frequency to derive throughput.
For DPF generation, we model the accelerator as a black-box generator parameterized by the throughput obtained from the synthesis. Rather than simulating internal microarchitectural details, we focus on system-level data movement and integration. Specifically, we model the generator as producing a stream of DPF outputs that are consumed by the evaluation pipeline including data movement and post-processing.

We implement this model using the GPU performance simulator \textit{Accel-Sim}~\cite{khairy2020accel}, which captures GPU kernel execution traces and replays them on a cycle-level simulator to model the detailed microarchitectural behavior of a target GPU design. The DPF generator is integrated into the memory hierarchy with input/output buffers placed close to the L2 cache partitions. 
\change{Our Accel-Sim model includes SM requests to DFA-mapped output buffers, the traversal of these requests through the GPU on-chip interconnect, and contention with normal memory traffic. The reported end-to-end results include the effect of DFA-targeted traffic sharing on-chip resources with ordinary GPU memory accesses.}

The following factors are explicitly modeled: \textbf{memory bandwidth contention} when large DPF data streams happen; \textbf{interconnect latency} across the GPU on-chip network and memory hierarchy; and \textbf{overlap} between DPF generation and evaluation (i.e., effectiveness of the generate-and-consume pipeline).
This methodology allows us to capture both the compute capability of DFA and the system-level bottlenecks from data movement and integration.

\textbf{Baselines.}
For private inference, we use state-of-the-art software implementations of FSS-based private inference as baselines (Orca~\cite{jawalkar2024orca} for CNNs, Sigma~\cite{gupta2023sigma} for Transformers). 
%
For PIR, we use a recent GPU-based DPF-PIR system~\cite{lam2023gpu} as a baseline.

\textbf{Workloads.}
%
For private inference, we ran ResNet-50~\cite{he2016deep}, BERT-Base~\cite{devlin2019bert}, GPT-Neo~\cite{brown2020language} and LLaMA2-7B~\cite{touvron2023llama} inference. Linear layers are evaluated using standard MPC primitives, while nonlinear layers (e.g., ReLU, comparison, softmax, and activation functions) are evaluated using FSS, which is accelerated with DFA in our proposed system. We record the per-sample latency number assuming LAN (1GBps, 0.05ms round trip time) and WAN (70MBps, 70ms round trip time) between the MPC parties, following prior work~\cite{huang2026beyond,gupta2023sigma,jawalkar2024orca}.
For PIR, we varied the database size between 16k--4M entries.
%
We considered batched query execution to capture realistic throughput-oriented deployments.

To estimate the private inference execution time with DFA, we measured the execution time on the baseline systems and replaced the execution time of FSS-related operations with the estimated time from the simulation.
Communication volume was measured as the total amount of data exchanged between parties per query, including both offline and online phases, where applicable. 
The volume of data movement captured the total amount of data transferred across system components, including CPU–GPU transfers and memory traffic, obtained from simulation and profiling. 
PIR execution time is directly from the simulation, as the workload is simple and can run entirely inside the simulator.
We estimated the GPU energy cost by measuring power through NVIDIA driver monitoring and combining it with the estimated latency. 


\subsection{Experimental Results}
\label{sec:results}

\subsubsection{Single-Path Workload: Private Inference}

For the system-level evaluation, we ran all the experiments on nodes equipped with an Intel Xeon Gold 6448Y CPU, 256GB main memory, and an A100 GPU. (Multiple nodes are used for private inference tasks, with network conditions controlled using Linux Traffic Control). We use the default environment setup provided by the baselines~\cite{gupta2023sigma,jawalkar2024orca, lam2023gpu}. In our experiment, we deployed 16 DPF units, each with 4 AES-128 engines (same for full-path workload experiments).

\autoref{tab:single_path_th} shows end-to-end performance. Here, the latency results do not include the disk reading latency since the baseline work does not model it, which we will discuss later. The results demonstrate that our design effectively eliminates the dominant bottlenecks in FSS-based inference. 
Offline communication is reduced from tens to hundreds of GBs to sub-GB. Trusted DFA also eliminates most of the key generation and storage overhead, significantly reducing disk traffic and memory pressure during the online phase. 
CPU-GPU communication is reduced by about $7.5\times$ to $62\times$, as large volumes of pre-generated keys no longer need to be transferred. These communication reductions directly translate into end-to-end speedups, with latency improving by more than $10\times$ under LAN. Under WAN, the improvement is still significant but relatively smaller, due to the online communication overhead contributing more to the overhead, which DFA does not optimize.
\change{
The reduced portion can be interpreted directly from the difference between rows: the offline-communication reduction is eliminated FSS key distribution, and the CPU--GPU-communication reduction is eliminated FSS/DPF key transfer and reread. 
}
\change{
We also include the ``GPU+Trusted DFA without generate-and-consume'' rows to demonstrate the importance of generate-and-consume optimization. This baseline removes offline key distribution but still store and rereads the generated FSS/DPF material (stored in and read from CPU DRAM; they are often too large to keep resident in GPU memory, especially for large-batch inference), so the remaining CPU--GPU traffic and latency show that extra overhead is brought to the execution time.
}

Energy consumption is also reduced, with improvements often exceeding $5\times$, reflecting both reduced data movement and more efficient execution.
When the hardware operates in the untrusted mode, energy savings is slightly smaller, since generation happens only at the trusted third party once, instead of twice at each party DFA. while the same computational workload is offloaded from the GPU to the DFA. However, the benefits are limited to computation efficiency; the end-to-end latency only sees minor improvements, and there is no reduction in network or CPU--GPU communication. This is because the protocol still relies on externally generated keys, and thus does not eliminate the data movement associated with the trusted third party, which is the dominant bottleneck.

\begin{figure}
    \centering
    \includegraphics[width=\linewidth]{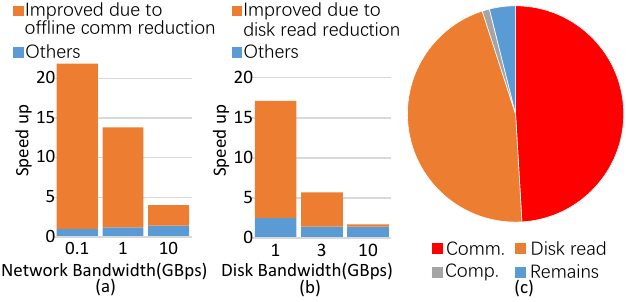}
    \caption{Breakdown of overhead reduction for BERT-base with trusted DFA under varying system conditions, with 1 GBps network and disk bandwidth when default: (a) Speedup decomposition across different network bandwidths for end-to-end execution with hidden disk reads. (b) Speedup decomposition across different disk read bandwidths during the online phase (without offline key transfer). (c) Under default setting, the proportion of execution time trusted DFA reduced through reducing communication, disk read (without disk read pipelining), computation overhead.}
    \label{fig:analyse}
\end{figure}

\textbf{Bottleneck Analysis.} The trusted DFA eliminates the need to transfer large volumes of FSS keys, significantly reducing offline communication, which is the key bottleneck. Although our accelerator also reduces the key evaluation overhead (less than 30\% of the overall overhead), computation is not the primary bottleneck for single-path workloads, and thus the improvement is limited. Also, key generation introduces non-trivial overhead in prior GPU-based designs. We show in~\autoref{fig:analyse} that the overall speedup is dominated by reductions in data movement rather than computation. Specifically, improvements in eliminating network transfer and disk I/O (when not hidden) account for the majority of the gains, while the contribution of reduced evaluation computation remains comparatively small.
Our generate-and-consume execution model removes this data transfer overhead entirely by avoiding disk-resident keys. Although increasing network and disk bandwidth can reduce the observed speedup, as discussed in prior work~\cite{huang2026beyond}, these costs remain significant in practice. In particular, transferring large volumes of FSS keys over WAN, a more realistic deployment setting than LAN, incurs substantial monetary cost regardless of bandwidth. Furthermore, storage overhead for pre-generated keys persists independently of disk read speed, making it a fundamental system-level concern.

\subsubsection{Full-Path Workload: PIR}


\autoref{tab:pir_results} shows the performance of PIR workloads. The results show that our accelerator significantly improves both throughput and energy efficiency across different database sizes. For a 16k-entry database, throughput increases by 8.52$\times$ and energy per request is reduced by 14.3$\times$. For a larger 1M-entry database, we still observe substantial gains, with throughput improving 6.31$\times$ and energy reducing by 10.5$\times$. 

The speedup is the smallest (5.1$\times$) with 1M entries when the GPU-DPF implementation~\cite{lam2023gpu} reaches its peak per-entry throughput at 1M entries. 
This is because the memory bandwidth and SM utilization are maximized at this point.
In contrast, our accelerator maintains a near-linear scaling trend with larger database sizes, as its performance is primarily determined by the throughput of the pipelined DPF engine. Consequently, our design achieves larger speedup with larger (4M+) entries.
Prior work~\cite{lam2023gpu} also showed that using different PRG, such as ChaCha20~\cite{bernstein2008chacha}, improves performance. 
We also saw a 1.5--2$\times$ speedup when using ChaCha20, but 
we mainly report AES-based numbers due to its wide popularity. 


\textbf{Bottleneck Analysis.} For full-path workloads, the bottleneck is dominated by computation for both the baseline and our DFA design. In the baseline, GPU utilization is consistently close to 100\% across all experiments. With a batch size of 512 queries, the observed GPU memory traffic remains relatively low (less than 200MBps for reads and 10MBps for writes), indicating that the workload is not memory-bandwidth bound but instead compute-bound.
For the DFA-equipped system, the computation bottleneck shifts to the DFA itself. The GPU performs only around 20GOPS of 128-bit integer multiplication, which is significantly below the A100’s peak capability (at the TOPS level), indicating that the GPU is underutilized. The theoretical peak throughput of 16 fully pipelined DPF units at 720MHz is approximately 350GBps, while Accel-Sim reports about 290GBps in practice. This indicates that the system operates close to its peak, with the remaining gap primarily due to interconnection and contention overhead.
\begin{table}[ht]
\centering
\caption{Full-path (PIR) workload performance comparison, with data entry size to be 128 bit. DFA does not need to be trusted for pure hardware acceleration.}
\vspace{-8pt}
\label{tab:pir_results}
\small
\begin{tabular}{l l c c}
\hline
\textbf{Database} & \textbf{Method} & \textbf{Throughput (k/s)} & \textbf{Energy/query (J)} \\
\hline
\multirow{3}{*}{16k entries}
 & GPU-DPF & 134 & 2.61 \\
 & GPU+DFA & 1125 & 0.187 \\
 & Ratio & 8.4$\times$ & 0.07$\times$ \\
\hline
\multirow{3}{*}{1M entries}
 & GPU-DPF & 3.48 & 100 \\
 & GPU+DFA & 17.58 & 11.9 \\
 & Ratio & 5.1$\times$ & 0.12$\times$ \\
\hline
\multirow{3}{*}{4M entries}
 & GPU-DPF & 0.788 & 443 \\
 & GPU+DFA & 4.39 & 47.7 \\
 & Ratio & 5.6$\times$ & 0.11$\times$ \\
\hline
\end{tabular}
\end{table}
\subsubsection{Hardware Cost Analysis}

\begin{table}[t]
\centering
\caption{Hardware cost comparison across technology nodes. We estimate the 7nm data using DeepScaleTool~\cite{sarangi2021deepscaletool}. The Cortex-M3-class control unit is modeled using parameter provided by existing commercial product~\cite{st33tpm12spi}.}
\vspace{-8pt}
\label{tab:tech_scaling}
\small
\begin{tabular}{l l c c c}
\hline
\textbf{Tech} & \textbf{Component} & \textbf{Timing} & \textbf{Area (mm$^2$)} & \textbf{Power (mW)} \\
\hline
\multirow{4}{*}{28nm}
 & 16 DPF units & 1.2 GHz & 1.73 & 29 \\
 & SRAM & 720 MHz & 3.06 & 585 \\
 & VPU & 1.04 GHz & 1.97 & 794 \\
 \cline{2-5}
 & DFA w/o Ctrl U & 720 MHz & 6.76 & 1408 \\
\hline
\multirow{4}{*}{7nm}
 & 16 DPF units & 1.5 GHz & 0.054 & 11 \\
 & SRAM & 910 MHz & 0.096 & 219 \\
 & VPU & 1.2 GHz & 0.062 & 298 \\
 \cline{2-5}
 & DFA w/o Ctrl U & 910 MHz & 0.212 & 528 \\
\hline
\hline
40nm & Ctrl Unit Core & 84 MHz & 0.040 & 0.92 \\
\hline
 28nm& SRAM & -- & 0.499 & 275.2 \\\cline{2-5}
 \multirow{2}{*}{28nm$^\dagger$}
 & ROM$^\dagger$ & -- & 0.021 & -- \\
 \cline{2-5}
 & Ctrl U Mem  & -- & 0.52 & 275.2 \\\hline
\hline
28nm & All DFA  & 720/96 MHz & 7.67 & 1683.5\\
\hline
7nm & All DFA  & 910/121 MHz & 0.239 & 631\\
\hline
\end{tabular}

\vspace{2pt}
\raggedright
\footnotesize
$^\dagger$ 
The Control unit core and memory size are collected from commercial documents~\cite{arm_cortex_m3,st33tpm12spi}. 
ROM is treated as cold storage and their active power is omitted. ROM area is (conservatively) estimated assuming $\sim$3.5$\times$ higher bit density than SRAM based on device-level comparisons~\cite{pentecost2021nvmexplorer}.
Timing of memory is not important for the low frequency core.
\end{table}

\begin{table}[ht]
\centering
\caption{Number of non-DPF 128-bit integer operations required for key generation per inference (in billions OPs).}
\vspace{-8pt}
\label{tab:vm_capacity}
\small
\begin{tabular}{l c c c}
\hline
\textbf{Workload} & \textbf{ADD} & \textbf{MUL} & \textbf{XOR} \\
\hline
ResNet-50 & 0.027 & 0 & 0.018 \\
BERT-Base & 0.037 & 0.012 & 0.007 \\
GPT-Neo & 0.320 & 0.064 & 0.138 \\
LLaMA2-7B &0.814 &0.114 &0.393 \\
\hline
\end{tabular}
\end{table}

We report the hardware cost of DFA in~\autoref{tab:tech_scaling}. Excluding the control unit, the accelerator occupies 6.76mm$^2$ and consumes 1.41W at 28nm, which scales to 0.212mm$^2$ and 528mW at 7nm. After incorporating the control unit and its associated memory subsystem, the total system cost increases to 7.67mm$^2$ and 1.68W at 28nm, and 0.239mm$^2$ and 631mW at 7nm.
The overhead is dominated by memory (64KB SRAM, 8KB ROM)
rather than the control core itself. 
The system is primarily bounded by the latency of the on-chip interconnect between SMs and L2 (modeled as $\sim$200 cycles in Accel-Sim~\cite{khairy2020accel}), which requires sufficient buffering to keep the DPF pipeline fully utilized. With a 20-cycle DPF pipeline, we provision buffers to sustain a batch size of 20.
Each DPF unit uses 128-bit dual-port SRAM with two output buffers, requiring $\sim$22KB per buffer in practice. For full-path workloads (e.g., PIR), internal buffering for traversal states dominates, requiring a 23KB double-buffered input buffer for key streaming.
These buffers are sufficient to hide input refill latency and sustain peak throughput. For single-path workloads, the buffering bottleneck is much smaller (sub-KB), keeping the trusted computation footprint small.
Within the accelerator, the DPF units account for only a small fraction of the total cost, while the memory and the programmable VPU dominate both the area and power. We also note that the SRAM frequency in our synthesis is conservative compared to modern designs, which limits the overall frequency of the system and suggests additional room for improvement. Overall, these results demonstrate that the proposed accelerator can be integrated with modest overheads.
We report the total non-DPF operations required per inference in~\autoref{tab:vm_capacity}. Although these operations grow with model size, they remain a small fraction of the overall computation compared to the DPF key generation. The 16-lane Ara design~\cite{cavalcante2019ara} we use for the VPU provides $\sim$33DP-GFLOPS throughput at 1GHz. In our simulation with the tools provided by Ara, this is sufficient to process these operations efficiently.



\subsubsection{Scalability Analysis}

We discussed the bottlenecks of our scheme and the baselines in the previous section. Here, we analyze how hardware scaling impacts the performance of our design. As noted earlier, the current bottleneck lies in the performance of the DFA. Scaling the hardware (e.g., doubling the number of DPF units and SRAMs) proportionally increases area and power and yields an approximately linear performance improvement in our experiments.
However, this scaling is ultimately bounded by system-level constraints. In particular, memory bandwidth imposes a hard upper limit. The NVIDIA A100 white paper reports an aggregate L2 read bandwidth of approximately 7TBps~\cite{nvidia2020ampere}, which serves as a practical limit for sustained data movement. 
To further push beyond this bottleneck, one potential design direction is to distribute DPF units closer to the SMs and route data directly to local output buffers, bypassing the NoC. While this approach may reduce interconnect pressure, it introduces new challenges, including inter-unit communication overhead (from the input buffer) and/or increased hardware complexity. These factors can offset the benefits by introducing additional pipeline stalls and coordination costs. As GPU architectures evolve, the maximum achievable capacity of our scheme primarily scales with improvements in NoC and memory bandwidth, rather than raw compute capability.

Our scheme scales linearly with the number of entries, matching the theoretical scaling behavior of pure GPU-based solutions. However, another important factor is the entry size. As the entry size increases, the complexity of the DPF evaluation remains unchanged, while the subsequent vector multiplication cost increases linearly. Specifically, each entry is partitioned into 128-bit chunks, and each chunk is multiplied by the expanded DPF output, increasing the overall computation.
This effect reduces the relative acceleration benefit of our design. Although the peak INT32 multiply throughput of the A100 is 19.5TOPS (corresponding to roughly 2TOPS for 128-bit integer multiplication), our current implementation only utilizes about 2.5GOPS, indicating that the system is strongly memory-bound under typical settings. However, as the entry size increases (e.g., beyond 16KB), the workload becomes increasingly dominated by vector multiplication. In this regime, the system shifts from being bandwidth-bound to compute-bound with respect to the DFA, and further scaling of DFA capacity no longer improves end-to-end throughput.

\section{Related Work}

Recent systems have significantly advanced the efficiency and practicality of FSS-based secure computation.
\textsc{AriaNN}~\cite{ryffel2020ariann} and \textsc{Pika}~\cite{wagh2022pika} first demonstrated private inference with FSS. The mixed-mode and fixed-point framework of~\cite{boyle2021function} extended FSS to more expressive numerical representations. At the same time, \textsc{Grotto}~\cite{storrier2023grotto} revisited the structure of DPF constructions to reduce both computation and communication by replacing multiple distributed comparison functions with more efficient primitives. More recent studies further improved system performance; \textsc{Sigma}~\cite{gupta2023sigma} and \textsc{Orca}~\cite{jawalkar2024orca} improved end-to-end secure inference efficiency by optimizing protocol composition and system-level execution, particularly targeting modern neural network workloads. Previous work has also explored the acceleration of MPC using untrusted hardware platforms such as FPGAs~\cite{wolfe2020secret, patel2020arithmetic, xu2025co}.

FSS/DPF has been widely used beyond private inference and PIR. These include private heavy-hitter protocols~\cite{boneh2021lightweight, mouris2024plasma}, private query systems that support expressive database operations~\cite{wang2017splinter}, and secret-shared storage systems such as Waldo and DORY~\cite{dauterman2022waldo, dauterman2020dory}. Similar techniques are also used in private function evaluation for RAM and ORAM-like settings~\cite{ji2023multi}, as well as in private set intersection protocols that can be viewed as batched DPF evaluations~\cite{garimella2024computation}. 
Across these systems, DPF serves as a core primitive for performing private index-based access or selection over large datasets, close to its functionality in PIR. 
Our work is complementary to these efforts and can be potentially applied to accelerate their underlying DPF generation/evaluation as well.

\textbf{MPC with Trusted hardware.}
PPMLAC~\cite{zhou2022ppmlac} and STAMP~\cite{huang2022efficient} proposed to accelerate private inference by offloading certain operations to a trusted programmable processor, thus reducing online latency and communication overhead. 
Our design supports both general single-path and full-path workloads, while they only accelerate fixed-functions targeting certain model inferences.
Also, our accelerator provides meaningful performance gains even when the accelerator is untrusted, preserving the full cryptographic security guarantees of the original FSS threat model.


Subsequent work explored similar directions using trusted execution environments (TEEs) or trusted hardware to accelerate secure computation. For example, prior studies leverage Intel SGX to accelerate cryptographic primitives such as bootstrapping and functional encryption~\cite{katz2007universally, lu2021correlated, fisch2017iron}, or to simplify secure computation protocols~\cite{choi2019hybrid, felsen2019secure}. Other approaches partition computation between MPC and trusted hardware, offloading selected operations to a TEE~\cite{gupta2016using, zhou2022ppmlac, wuhybrid}.
While these works share the high-level goal,
they typically rely on relatively powerful TEEs. 
\section{Conclusion}

In this paper, we present DFA, a hardware--software co-designed system that accelerates FSS by targeting its dominant primitive, DPF. DFA addresses the main system bottlenecks---key generation, storage, transfer, and data movement---through a DPF-centric accelerator and a generate-and-consume execution model. Across private inference and PIR workloads, DFA achieves over 10$\times$ latency reduction, more than 20$\times$ communication reduction, over 5$\times$ energy savings for private inference, and over 5$\times$ throughput improvement with around 10$\times$ energy savings for PIR.

\section*{Acknowledgments}
This work is partly supported by the U.S. National Science Foundation under award No. CCF-2118709 and CCF-2529883. Any opinions, findings, and conclusions or recommendations expressed in this material are those of the author(s) and do not necessarily reflect the views of the National Science Foundation. 

\bibliographystyle{ACM-Reference-Format}
\bibliography{BIBFILE}

\end{document}